\documentclass{article}
\usepackage{epsfig}

\begin{document}

\title{Robust control in the quantum domain}

\author{Andrew Doherty$^{1}$, John Doyle$^{1}$, Hideo Mabuchi$^{1}$,\\
Kurt Jacobs$^{2} $, Salman Habib$^{2}$ \\ $^{1}$California
Institute of Technology, $^{2}$Los Alamos National Laboratory}
\maketitle
\thispagestyle{empty}\pagestyle{empty}

\begin{abstract}
Recent progress in quantum physics has made it possible to perform
experiments in which individual quantum systems are monitored and manipulated
in real time. The advent of such new technical capabilities provides strong
motivation for the development of theoretical and experimental methodologies
for {\em quantum feedback control}. The availability of such methods would
enable radically new approaches to experimental physics in the quantum realm.
Likewise, the investigation of quantum feedback control will introduce
crucial new considerations to control theory, such as the uniquely quantum
phenomena of entanglement and measurement back-action. The extension of
established {\em analysis} techniques from control theory into the quantum
domain may also provide new insight into the dynamics of complex quantum
systems. We anticipate that the successful formulation of an input-output
approach to the analysis and reduction of large quantum systems could have
very general applications in non-equilibrium quantum statistical mechanics
and in the nascent field of quantum information theory.
\end{abstract}

\section{Introduction}

It would be of great interest in quantum physics to develop controlled and
systematic methods for deriving approximate descriptions of complex dynamical
systems. A range of powerful techniques that may be applicable to quantum
scenarios have been previously investigated in the context of robust control
theory; here we are interested in model reduction via balanced truncation in
the Hankel norm \cite{zhou}. Balanced truncation and related methods
(\textit{e.g.} based on the gap metric) are particularly attractive due to
the availability of tight error bounds and for giving considerable insight
into the interconnection of dynamical systems. Tractable schemes for quantum
model reduction would have important applications in non-equilibrium
statistical mechanics and in the development of quantum information
technology -- one of our main long-term goals will be to use model reduction
to facilitate numerical simulation of fault-tolerant architectures for
quantum computers \cite{preskill}. This conference paper describes our first
results on balanced truncation of coupled quantum systems and the
formulation of the input-output descriptions of the dynamics of
quantum error-correcting codes.

We anticipate that a general program of extending modern control-theoretic
methods to quantum scenarios would have tremendous impact on experimental
research in quantum physics as well \cite{cqed}. In this context we are
particularly interested in robust controller synthesis and controller
reduction methodologies for feedback control of open quantum systems via
real-time processing of measured output signals. New experimental
techniques in the manipulation of quantum systems show the potential for
genuinely quantum technologies such as quantum computers, however
active feedback and control will be crucial to their functioning and
motivate the adoption of control theory concepts and methods in
quantum physics generally. 

\section{Quantum Dynamics and Linear Systems Theory}

The first step in applying control-theoretic approaches to the approximation
of quantum dynamics is to write the Schr\"{o}dinger equation as a system of
linear ordinary differential equations for an appropriate set of dynamical
variables that characterize the quantum state. This set of variables will
depend on the physical system, and more specifically on the parameters that
are of most interest in a particular problem. One general approach in the
spirit of the Shr\"{o}dinger picture of quantum dynamics is to write the
first order linear differential equations for the matrix elements of the
quantum density operator and to separate those out into quantities of
interest and quantities to be reduced. An approach more natural from the
Heisenberg picture is to write the dynamics in terms of linear ordinary
differential equations for operator expectation values -- as long as a quorum
of system observables is chosen this is precisely equivalent. A common
example of such a quorum is the elements of the Bloch vector $(\langle
X\rangle ,\langle Y\rangle ,\langle Z\rangle),$ that
completely define the quantum state of a two-state system. The
operators $X,Y,Z$ are the familiar Pauli operators. 

It is often the case that only the reduced state of some subsystem is of
interest. An important question is to what extent a simple description of the
dynamics (`subdynamics') of this reduced system may be found. Here we
consider the specific example of two coupled two-state systems (`spins'),
where only the state of the first spin is of direct interest, and where the
second spin is phase damped by coupling to an infinite reservoir. Although
this is an almost trivial example and full numerical integration of the
dynamics would pose no computational difficulty for such a low-dimensional
problem, the procedures we follow are simple and systematic and may thus be
applied to a wide range of more complex quantum systems of significant
physical interest.

In our example, the state of the full system is described by the Bloch
vectors of the two spins ($\langle I_{1}\rangle ,\langle J
_{2}\rangle ,I,J=X,Y,Z$) along with {\em nine} other quantities of the form
$\langle I_{1} J_{2}\rangle$. The necessity of having so many
`interconnection variables' in this problem immediately exemplifies one of
the fundamental differences found in interconnecting quantum systems -- the
state space associated with the joint system is very much larger than would
be the case for classical dynamics, as a result of the tensor product
structure of the Hilbert space. Rather than having $3N$ degrees of freedom
(where 3 is the number of variables describing an isolated single spin
system), a system made up of $N$ interacting spins will have $2^{N}-1$
degrees of freedom. The multitude of extra parameters characterizes the
entanglement (quantum correlation) of the individual systems. (We recognize
that the same dimensional arithmetic holds for the dynamics of classical
probability {\em distributions}, but the quantum feature here is that no
underlying `trajectory' picture exists in which the state spaces may be
combined by direct sum.) Hence, we expect model reduction techniques to be of
great utility in the general study of interconnected quantum systems.

The quantities of interest in our example are a quorum of expectation values
for a physical subsystem, namely the first spin. This partitioning of the
variables according to physical subsystems is not the only type of problem
that may be considered; it may be the case for example that the full system
is simply one high-dimensional system for which only a few expectation values
are of interest, with other moments being important only as a means of
calculating these. In any such scenario it is possible to proceed exactly as
we do here. Another important consideration is that even for a system made up
of many coupled subsystems it may not be the reduced density matrix for any
given subsystem that is of primary interest. In quantum error-correcting
codes, for example, it is the Bloch vector corresponding to the state of the
logical (encoded) spin that is of interest, which is embedded within a highly
entangled subspace of a string of real (physical) spins. Approximate
simulations of realistic (imperfect) error correction protocols may well be
facilitated by exactly the means described here, with an appropriate
parametrization of the joint system state as is discussed below. In
general, the selection of a 
particular set of parameters with which to describe the dynamics of the
overall system must be based not only on knowledge of the dynamics and the
Hilbert space structure of the system but also by what features of these
dynamics are relevant to the question at hand.

Getting back to our example, we imagine that the two systems have a simple
coupling and as we mentioned above the second spin suffers phase damping. If
we did not introduce any dissipation the eigenvalues associated with our
system of linear differential equations would all have real parts equal to
zero (unitary dynamics). By adding the phase damping we guarantee that both
the overall system and the subsystem on which the model reduction is
performed are Hurwitz. Physically we are in any case most interested in
applications to open quantum systems where there will almost always be some
dissipative dynamics which usually ensures the stability of the
equations of motion. 
Thus we consider the following master equation for the overall system,
\begin{eqnarray}
\dot{\rho} &=&-i[H,\rho ]+\gamma \mathcal{D}[Z_{2}]\rho , \\
H &=&\frac{1}{2}\hbar \omega _{1}Z_{1}+\frac{1}{2}\hbar \omega
_{2}Z_{2}+\hbar g X_1 X_2,
\end{eqnarray}
where for an arbitrary operator $c$, $\mathcal{D}[c]\rho =c\rho c^{\dagger }-
\frac{1}{2}c^{\dagger }c\rho -\frac{1}{2}\rho c^{\dagger }c$. This operator
differential equation can be converted to an equivalent set of ordinary
differential equations for a set of operator expectation values by standard
techniques \cite{walls}. Here we write only those equations that couple to
the Bloch vector of the first spin:
\begin{eqnarray}
\langle \dot{X}_{1}\rangle  &=&-\omega _{1}\langle Y_{1}\rangle
,  \label{exps} \\ \langle \dot{Y}_{1}\rangle  &=&\omega _{1}\langle
X_{1}\rangle -2g\langle Z_{1}X_{2}\rangle ,  \nonumber
\\
\langle \dot{Z}_{1}\rangle  &=&2g\langle Y_{1}X_{2}\rangle
,  \nonumber \\ \langle \dot{Z_{1} X}_{2}\rangle
&=&-2\gamma\langle Z_{1}X_{2}\rangle -\omega _{2}\langle Z_2
Y_{2}\rangle +2g\langle Y_{1}\rangle ,  \nonumber
\\
\langle \dot{Z_{1} Y}_{2}\rangle  &=&\omega _{2}\langle Z_2
X_{2}\rangle -2\gamma \langle Z_{1}Y_{2}\rangle , 
\nonumber \\ \langle \dot{X_{1}X}_{2}\rangle  &=&-2\gamma
\langle X_{1}X_{2}\rangle -\omega _{1}\langle Y_{1}
X_{2}\rangle 
-\omega _{2}\langle X_{1}Y_{2}\rangle
,  \nonumber \\ \langle \dot{Y_{1}X}_{2}\rangle  &=&-2g\langle
Z_{1}\rangle +\omega _{1}\langle X_{1}X_{2}\rangle
-2\gamma \langle Y_{1}X_{2}\rangle \nonumber \\
 && -\omega _{2}\langle Y_{1}
Y_{2}\rangle ,  \nonumber \\ \langle \dot{X_{1}
Y}_{2}\rangle  &=&\omega _{2}\langle X_{1} X_{2}
\rangle -2\gamma \langle X_{1}Y_{2}\rangle -\omega
_{1}\langle Y_{1}Y_{2}\rangle \nonumber  \\
&& -2g\langle Z_2\rangle ,
\nonumber \\ \langle \dot{Y_{1}Y}_{2}\rangle  &=&\omega
_{2}\langle Y_{1}X_{2}\rangle +\omega _{1}\langle X_{1}
Y_{2}\rangle -2\gamma \langle Y_{1}Y_{2}\rangle ,
\nonumber \\ \langle \dot{Z}_{2}\rangle  &=&2g\langle
X_{1} Y_{2}\rangle .  \nonumber
\end{eqnarray}

Having formed a useful description of the dynamics of the state in terms of a
set of real parameters, ${\bf x}$, and identified the parameters of interest,
${\bf x}_{1}$, and those which are not, ${\bf x}_{2}$, it is straightforward
to recast our problem in the input-output formalism of control theory. The
system of equations above is of the form
\begin{equation}
{\bf \dot{x}}=A{\bf x,}
\end{equation}
but may always be written in the form
\begin{eqnarray}
{\bf \dot{x}}_{1} &=&A_{1}{\bf x}_{1}+B_{1}{\bf y}_{2}, \\ {\bf \dot{x}}_{2}
&=&A_{2}{\bf x}_{2}+B_{2}{\bf y}_{1}, \\ {\bf y}_{1} &=&C_{1}{\bf
x}_{1},\quad {\bf y}_{2}=C_{2}{\bf x}_{2}.
\end{eqnarray}
This explicitly formulates the full dynamics of the quantum system in terms
of inputs and outputs of two systems, one of which of is of physical
interest, and another for which we want to find a low-order approximation.
These two systems are connected in such a way that the outputs of one system
are the inputs of the other. The terms arising from the unitary interaction
of the two subsystems guarantee that any quantum system will have this kind
of structure where the system of interest both drives and is driven by the
environment. Only in certain extreme limits where the `environment' (the set
of degrees of freedom other than the subsystem of interest) is essentially
infinite-dimensional is it possible to formulate an approximate treatment for
which the transfer function from inputs to the environment to outputs from
the environment is essentially zero (for any finite time-horizon) and thus
the outputs from the system of interest do not affect its subsequent
evolution. In physics this is known as the Markov limit, and is commonly
invoked in the derivation of quantum master equations \cite{walls}.

Returning to our example, the system $(A_{2},B_{2},C_{2})$ is both observable
and controllable so it is immediately possible to find a similarity
transformation for the state space that results in a balanced realization of
this state space model. That is, there exists some matrix transformation $T$
that results in a transformed state space realization ${\bf
\tilde{x}}_{2}=T{\bf x}_{2},\tilde{A}_{2}=TA_{2}T^{-1},\tilde{B}
_{2}=TB_{2},\tilde{C}_{2}=C_{2}T^{-1}$ such that the associated
controllability and observability gramians are equivalent and diagonal. Their
diagonal elements are the Hankel singular values. This allows a balanced
truncation of the model in which dimensions of the transformed state space
that correspond to small Hankel singular values are simply disregarded. The
resulting state space realization is also balanced and stable so long as the
truncation is performed such that the smallest Hankel singular value included
in the reduced state space model $(\bar{A}_{2},\bar{B}_{2},\bar{C}_{2})$ is
greater than the largest one that is excluded.

The error in this approximation may be tightly estimated since the transfer
function corresponding to the new reduced state space model is close to the
original transfer function, in the sense that the $H_{\infty}$ norm of their
difference is bounded from above by twice the sum of the disregarded Hankel
singular values and from below by the largest disregarded singular value.
Given that our example seems to have the general flavor of a controller
reduction problem, we suspect that more sophisticated techniques based on gap
metric or structured singular value may be applicable \cite{zhou}. However,
we find that the simple balanced truncation used here is more than sufficient
to find accurate approximations to the overall dynamics of our example model.

To take a concrete example, imagine that in our example system we are only
concerned with the inversion of the first spin $\langle Z_{1}\rangle$.
This means that we are interested in the dynamics of one of the 15 parameters
defining the quantum state. The system of equations (\ref{exps}) shows that
$\langle Z_{1}\rangle $ is coupled to only 5 of these parameters. The
others may be ignored, as a simple consequence of the specific coupling we
have chosen. Moreover, numerical computation of balanced realizations shows
that in the broad parameter regime $g>\gamma >\omega _{1},\omega _{2}$ only
one of these parameters has a significant effect on the dynamics and so it is
possible to find an approximate description of the dynamics of the
environment with a state space that is only one dimensional by truncating
this balanced realization. This parameter regime corresponds to the situation
in which although the first spin is strongly damped the decay is very
non-Markovian and so a naive adiabatic elimination is of no use (see Figure).

 \begin{figure}[ht]
\epsfxsize=2.5in \epsfclipon \centerline{\epsffile{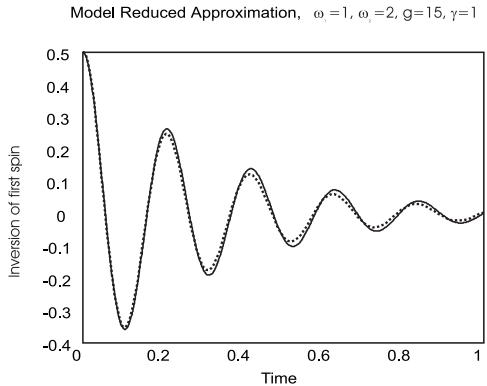}}
\caption{Numerical test of
model reduction via balanced truncation for a simple quantum system: solid
curve is derived from a numerical integration of the dynamical equations
without approximation, dotted curve is derived from numerical integration of
the reduced model.}
\end{figure}

Two advantages of this technique bear emphasizing. Firstly, the approximation
is controlled in the sense that it is possible to obtain a rigorous estimate
of the error resulting from truncating the balanced realization. Thus it is
possible to see in advance whether a particular approximation is well
justified. In other areas of the parameter space of this model the dynamics
are more complicated and the radical approximation given above is not at all
accurate. However, the boundaries of the simple regime are indicated by
Hankel singular values which are larger and more nearly equal.

A second advantage of this method is that unlike more common (physical)
approaches to modeling quantum subdynamics, there is no difficulty in dealing
with entangled initial states of the system and environment. Corresponding to
any initial state of the system, including all entangled states, there is
some initial value of ${\bf x}_{1}$ and ${\bf x}_{2}$. The transformation to
the balanced realization determines the appropriate initial state ${\bf
\tilde{x}}_{2}=T{\bf x}_{2}$ in the state space of the balanced truncation.
It makes no difference whether these vectors correspond to an entangled state
or a product state of the two spins. This flexibility is in marked contrast
to the situation for adiabatic elimination where it is usually necessary to
assume that the system and the environment are in a product state.

\section{Applications to Quantum Error Correction}
A significant potential application of this approach to the approximate
simulation of quantum systems is in the area of quantum error
correction for a quantum computer. In a quantum computation quantum
mechanical two-state systems or qubits replace bits as the fundamental
means of storing and manipulating information. The main challenge in
building a quantum computer, or simply an accurate quantum memory, is
in achieving sufficient control over the state of a quantum system
that it is possible to accurately perform the computation and avoid
error processes. Quantum error correction provides a possible solution
to the problem of random errors caused by uncontrolled couplings of
the computer to its surrounding enviroment. In quantum error 
correction (see~\cite{gottesman} for a technical introduction), as in 
its classical counterpart, the information stored in memory is
protected from errors by encoding logical qubits in redundant physical
degrees of freedom --- these will typically be two-state
systems such as those discussed above and we sometimes refer to them
as physical qubits in the
following. 

The theory of fault-tolerant quantum
computation has shown that by using encoded operations on the
logical qubits and by encoding the logical qubits in many layers of coded
states (concatenating codes) it is possible to perform an arbitrary
computation with any desired level of accuracy given that the
fundamental error rate is below some bound. Estimates exist for the
value of the bound for at least some choices of code,
fault-tolerant gate set and error process. It is still of
interest to be able to simulate the performance of error-correction
and fault-tolerant computation in the presence of errors which are not
accounted for by the code  and also to find tighter
bounds on the fundamental error rates necessary to achieve
fault-tolerance. However, while it has been possible to simulate the
evolution of simple three and five qubit codes, the full
simulation of concatenated codes or fault tolerant computing schemes
is not currently possible and would appear to be very challenging as a
result of the exponential growth of the state space as the number of
physical qubits increases. In a simple quantum code a qubit is encoded
in five physical qubits, each level of concatenation will then use five
qubits from the previous level in order to encode the qubit at the
next highest level leading to a state space of $2^{5^{N}}$ dimensions
to encode a single logical qubit, although the reduced state of the
logical qubit is
described by just three parameters.
This is an example of a quantum mechanical system where only a
relatively small amount of information about the state is of interest
(the reduced state of the logical qubit) but this subsystem has
non-trivial interaction with the very large state space. The
application of our state-space based approach to helps firstly
to identify the variables in the state space that affect the state of
the logical qubit leading to a very great simplification of the model
and secondly provides an approach to determining the 
approximate evolution of a concatenated code under various physically
reasonably error processes using the kind of model reduction described
above. Here we confine ourselves to describing the fundamental
building blocks of the problem leaving a full treatment to further
work.

In any eventual quantum computer it will be necessary to
minimize couplings to the environment. These will lead to errors of the
general form  
\begin{equation}
\rho \rightarrow \sum_i E_i \rho E_i^{\dagger},
\end{equation}
where, for a quantum memory, error processes correspond to $E_i$
 different from the identity. For example, $E_i=X_{1}$ would
 flip the sign of the first qubit in the quantum memory by swapping
 the probability 
 amplitudes in each of its two levels --- we refer to this as a bit
 flip error. 
 In a quantum error correcting scheme a logical qubit is encoded in the
state of several physical qubits, the space of states of the logical
qubit is then a subspace of the complete Hilbert space. The code is
arranged such that the dominant error processes (for example bit
 flips) take the 
physical state into orthogonal subspaces of the total Hilbert
 space. Recovering from the errors then requires making a projective 
measurement onto these orthogonal subspaces to determine which error
occured (syndrome identification) and then an appropriate recovery
operation to rotate the state back into the computational subspace. The
possible outcomes of this measurement correspond to projection
operators $P_{i}$ onto the code and error subspaces and we will label
the corresponding recovery operators $R_{i}$. The expected state of
the system after the recovery operation is then 
\begin{equation}
\bar{\rho} =\sum_{i}R_{i}P_{i}\rho \left( t\right) P_{i}R_{i}.
\end{equation}
Presuming that these operations are perfect this state is now on the
code subspace and so it may be characterized by the expectation values of
the corresponding logical qubit $\langle \bar{X}\rangle,\langle
\bar{Y}\rangle,\langle \bar{Z}\rangle$. These correspond to particular
linear combinations of the expectation values characterizing the state
of the physical qubits prior to the recovery operations. 

We will consider a simple three
qubit code which corrects bit-flip errors. The simplest
code which corrects for all independent errors on the individual
physical systems requires five physical qubits. For the bit flip code
projective measurements are made of the observables $Z_1Z_2$ and
$Z_1Z_3$ and the four outcomes indicate either that no error has
occured or that the state of one of the physical qubits has been
flipped and if so which qubit has changed. As a result the recovery
operators are $X_1,X_2,X_3$. It is then possible to write the
expectation values of the logical qubit after recovery in terms of the
physical expectation values prior to recovery
\begin{eqnarray*}
\langle \bar{X}\rangle  &=&\langle X_{1}X_{2}X_{3}\rangle  \\
\langle \bar{Y}\rangle  &=&\frac{1}{2}\left( \langle X_{1}X_{2}Y_{3}\rangle
+\langle Y_{1}Y_{2}Y_{3}\rangle +\langle X_{1}Y_{2}X_{3}\rangle
\right.
\\ && \left.   +\langle
Y_{1}X_{2}X_{3}\rangle \right)  \\
\langle \bar{Z}\rangle  &=&\frac{1}{2}\left( \langle Z_{1}\rangle +\langle
Z_{2}\rangle +\langle Z_{3}\rangle -\langle Z_{1}Z_{2}Z_{3}\rangle \right) .
\end{eqnarray*}
This particular transformation assumes that the measurement and
recovery steps are perfectly realized. This is by no means necessary,
errors in either step would simply lead to a different dependence of
the logical qubit expectation values on the pre-recovery expectation
values. Particularly easy to take account of would be a noisy
projection which would replace the projectors $(I \pm Z_1Z_2)/2,(I\pm
Z_1Z_3)/2$ with $(I\pm \eta Z_1Z_2)/2,(I\pm \eta Z_1Z_3)/2$ where $\eta$
indicates the efficiency of the measurement and with $\eta=0$ the
measurement results would be completely random. Similarly a noisy
recovery could be simulated by replacing $\rho \rightarrow X_i\rho X_i$
by  $\rho \rightarrow \eta X_i\rho X_i + (1-\eta)\rho$ as the recovery
operation. These simple
choices do not change the specific physical expectation values
involved in the above equation just their coefficients, however, more
general noise models introduce a dependence on more expectation
values. In this way it is possible to identify the expectation values
on the total Hilbert space which are relevant to evolution of the
logical qubit given a specific measurement and recovery process.

The particular error model for the time between recovery steps
determines the dynamics of the 
physical expectation values such as $\langle
X_{1}X_{2}X_{3}\rangle$. An example of such an error model is the
Lindblad master 
equation describing the possibility of random bit flips
\begin{eqnarray*}
\dot{\rho} &=&\Gamma \mathcal{D}[X_{1}]\rho +\Gamma \mathcal{D}[X_{2}]\rho
+\Gamma \mathcal{D}[X_{3}]\rho  \\
&=&\Gamma \left( X_{1}\rho X_{1}+X_{2}\rho X_{2}+X_{3}\rho X_{3}-3\rho
\right) .
\end{eqnarray*} 
This describes a situation where the physical qubits undergo
independent bit flips at a constant rate.  Since the code corrects
for a single bit-flip, the state of the logical 
qubit after the correction should be unaffected by the noise to first order in
time. Such a master equation could
arise in the physical description of qubits subject to noisy 
magnetic fields affecting each of the qubits independently. 
In order to see how the code functions under this noise model, lets
consider then the the evolution of $\langle 
\bar{Z}(t)\rangle$ (this is an expectation value of the logical qubit
given that a recovery operation is performed at time $t$). The logical qubit
expectation value after recovery may be related to the physical expectation
values before recovery by 
\begin{equation}
\left(
\begin{array}{c}
\langle \bar{Z}\rangle  \\ 
\alpha 
\end{array}
\right) = \left(
\begin{array}{cc}
\frac{1}{2} & -\frac{1}{2} \\ 
\frac{1}{2} & \frac{1}{2}
\end{array}
\right) \left(
\begin{array}{c}
Z_T  \\ 
\langle Z_{1}Z_{2}Z_{3}\rangle 
\end{array}
\right),
\end{equation}
where $\langle Z_T\rangle =\langle Z_{1}\rangle +\langle Z_{2}\rangle +\langle
Z_{3}\rangle$. The auxiliary degree of freedom $\alpha$ is a linear
combination of the physical qubit expectation values that interacts
with $\langle \bar{Z}\rangle$. None of the other degrees of freedom in
the problem have an effect. The
chosen error model determines the evolution of the physical qubits
\begin{equation}
\frac{d}{dt}
\left(
\begin{array}{c}
Z_T \\ 
\langle Z_{1}Z_{2}Z_{3}\rangle .
\end{array}
\right) = \left(
\begin{array}{cc}
-2\Gamma & 0 \\ 
0 & -6\Gamma
\end{array}
\right) \left(
\begin{array}{c}
Z_T \\ 
\langle Z_{1}Z_{2}Z_{3}\rangle 
\end{array}
\right).
\end{equation}
As a result it is straightforward to determine the time evolution
of $\langle \bar{Z}(t)\rangle$ and $\alpha$
\begin{equation}
\frac{d}{dt}\left(
\begin{array}{c}
\langle \bar{Z}\rangle  \\ 
\alpha 
\end{array}
\right)=\left(
\begin{array}{cc}
-2\Gamma & \Gamma \\ 
\Gamma & -2\Gamma
\end{array}
\right)\left(
\begin{array}{c}
\langle \bar{Z}\rangle  \\ 
\alpha 
\end{array}
\right).
\end{equation}
If at $t=0$ the state of the system is in the code subspace with
$\langle \bar{Z}\rangle=Z_0$ then 
\begin{equation}
\langle \bar{Z} \rangle (t)= \frac{1}{2} Z_0 (3e^{-2\Gamma
  t}-e^{-6\Gamma t}).
\end{equation}
As claimed, the logical qubit expectation value changes only at second
order in $t$.  This is true of the other expectation values also since
the 
error model chosen here does not lead to time dependence of $\langle
X_{1}X_{2}X_{3}\rangle $ and the equations for
$\langle \bar{Y}\rangle$ result in the same expressions as for $\langle
\bar{Z}\rangle=Z_0$. So we have developed here an input-output picture
exactly similar to our previous discussion. Degrees of freedom
describing the state of the qubit after an error correction cycle
interact with a relatively small number of degrees of freedom that
describe the effects of noise. Here there is only one such
`environment' variable mediating the effects of this simple noise
process on the logical qubit of our particular code, but more
sophisticated codes and error models will certainly lead to more
complicated systems. Such a formulation of error correction
is a necessary precursor the development of simple descriptions of the
evolution amenable to model reductions of the kind discussed above. In
particular we are interested in applying these techniques to
concatenated codes and eventually fault tolerant computation. 

One of the main interests of this approach is to consider the effect
of errors other than those for which the code is most
effective. Having determined the transformations describing
syndrome detection and error recovery it is straightforward to
implement the time dependence of the physical expectation values
appropriate for any given error model and thereby derive the appropriate
input-output model for the evolution of the logical qubit. One
example which would be likely to be of interest in any physical
implementation of the bit flip code is the situation where the errors
on each physical qubit are correlated. The Lindblad master equation 
\begin{equation}
\dot{\rho}=\Gamma_c \mathcal{D}[X_{1}+X_{2}+X_{3}]\rho 
\end{equation}
describes the effect of a noisy `stray' magnetic field which affects
all three of the qubits. In this case the equations for the time
evolution of $\langle Z_{1}Z_{2}Z_{3}\rangle $ and $\langle
Y_{1}Y_{2}Y_{3}\rangle $ couple to other expectation values rather
than simply damping as they do in the original model. As in the
first error model the differential equations for the logical qubit 
expectation values couple to variables describing the state of the
environment --- in this case it turns out that there are two such 
degrees of freedom and the resulting
evolution of the logical qubit has a first order time dependence
proportional to $\Gamma_c$.  

In concatenated coding there are several levels of error
correction. The logical qubits described above are in turn combined into
triples which code for a logical qubit at a higher level.  The
measurement and correction transformations are applied 
firstly to the triples of physical qubits at the lowest level exactly as
described above and then at each
higher level of the code. Thus for a single concatenation of the above
code there are nine physical qubits and the overall tranformation
between the top level logical qubit and the physical qubits may be
derived by applying the previous formula to each level of the code and
is of the form
\begin{eqnarray}
\langle \bar{Z}\rangle&=& \frac{1}{4}\Biggl( \sum_{i=1}^{9}\langle Z_{i}\rangle
-\sum_{i,j,k} c^{(3)}_{ijk}\langle Z_{i}Z_{j}Z_{k}\rangle
\Biggr. \nonumber \\
&& + \sum_{i,j,k,l,m} c^{(5)}_{ijklm}\langle
    Z_{i}Z_{j}Z_{k}Z_{l}Z_{m}\rangle 
 \nonumber \\
&&  - \sum_{i,j,k,l,m,n,o} c^{(7)}_{ijklmno}\langle
Z_{i}Z_{j}Z_{k}Z_{l}Z_{m}Z_{n}Z_{o}\rangle 
 \nonumber \\
&& \Biggl. 
+\langle
Z_{1}Z_{2}Z_{3}Z_{4}Z_{5}Z_{6}Z_{7}Z_{8}Z_{9}\rangle \Biggr) ,
\end{eqnarray}
for a set of coefficients $c^{(i)}$.
Each of the five terms in the above sum decays at a given rate under
the first error model discussed above and this means that it is
necessary to 
keep track of only five degrees of freedom in this concatenated code
in order to determine the evolution of $\langle \bar{Z}\rangle$ --- the
value of $\langle \bar{Z}\rangle$ itself and four degrees of freedom
that mediate the effects of noise on the logical qubit. So there are
only eight degrees of freedom (including those necessary to
solve for $\langle \bar{Y}\rangle$ ) that are in fact coupled to the
state of the logical qubit 
even though there are 511 degrees of freedom in the full quantum state. 
The scaling in the size of the problem for this particular error model is $3^n$
where $n$ is the number of levels of concatenation which is more
favourable than the scaling of the entire Hilbert space $2^{3^n}$.

In general the protection of a quantum memory will consist of
periods where the quantum state is left to decohere followed by the
application of a possibly noisy error detection and
correction. As in the previous section we
have formulated the description of this evolution in terms of a
logical qubit with an input-output coupling to a certain state-space
model describing the effects of noise on the logical qubit. The
resulting systems are analytically simple for individual codes and
simple error models, however the dimensionality of the problem grows
with levels of concatenation of the code and we anticipate that model
reduction techniques will be useful in obtaining tractable models for
the overall evolution.
So far we have discussed the development of 
description of a single
decoherence-recovery iteration. This overall transformation, once
determined, is  
applied repeatedly to 
describe the build-up of errors in the quantum memory. This is a
discrete time model which may itself be amenable to some form of
model reduction. It remains for further work however to develop
parallel techniques for describing the interaction of logical
qubits during a computation and thus address the question of fault
tolerant computation itself. 

\section{Conclusions}

In this paper we have briefly outlined the initial stages in a program
of applying the analysis techniques of robust control theory to
quantum systems. We believe that these techniques will have broad
application and in particular will enable the
simulation of the dynamics of a quantum mechanical memory that employs
a highly concatenated quantum error correcting code and perhaps also
of a fault-tolerant quantum computation. The application of both
analysis and synthesis techniques taken from robust control theory
will play an important role in the development of emerging quantum
technologies, such as quantum computation.

\end{document}